%% file: overlap.tex
\documentclass[preprint,aps,tightenlines,floatfix,showpacs,byrevtex,nofootinbib, superscriptaddress]{revtex4}
\usepackage{pictex}
\usepackage{graphics}
\usepackage[dvips]{graphicx}

\include{commands}

\def\be{\begin{eqnarray}}
\def\ee{\end{eqnarray}}

\def\lb{\lbrack}
\def\rb{\rbrack}

 \def\Slash#1{
  \begin{picture}(5,6)(0,0)
  \put(-.7,-1.2){\line(5,6)6}
  \end{picture}
  \kern-.8em#1}
 \def\slash#1{
  \begin{picture}(5,6)(0,0)
  \put(-1.5,-1.7){\line(5,6)5}
  \end{picture}
  \kern-.8em#1}

\def\Sn{\Slash \nabla}
\def\sd{\Slash \partial}

\begin{document}
\preprint{ADP-01-42/T474}

\title{Accelerated Overlap Fermions}

\author{Waseem Kamleh}
\email{wkamleh@physics.adelaide.edu.au}
\affiliation{CSSM Lattice Collaboration, Special Research Centre for the Subatomic Structure of Matter and Department of Physics and Mathematical Physics, University of Adelaide 5005, Australia. }
\author{David H. Adams}
\email{adams@phy.duke.edu}
\affiliation{CSSM Lattice Collaboration, Special Research Centre for the Subatomic Structure of Matter and Department of Physics and Mathematical Physics, University of Adelaide 5005, Australia. }
\affiliation{Physics Dept., Duke University, Durham, NC 27708, USA }
\author{Derek B. Leinweber}
\email{dleinweb@physics.adelaide.edu.au}
\affiliation{CSSM Lattice Collaboration, Special Research Centre for the Subatomic Structure of Matter and Department of Physics and Mathematical Physics, University of Adelaide 5005, Australia. }
\author{Anthony G. Williams}
\email{awilliam@physics.adelaide.edu.au}
\affiliation{CSSM Lattice Collaboration, Special Research Centre for the Subatomic Structure of Matter and Department of Physics and Mathematical Physics, University of Adelaide 5005, Australia. }

\begin{abstract}
Numerical evaluation of the overlap Dirac operator is difficult since it contains the sign function $\epsilon(H_{\rm w})$ of the Hermitian Wilson-Dirac operator $H_{\rm w}$ with a negative mass term. The problems are due to $H_{\rm w}$ having very small eigenvalues on the equilibrium background configurations generated in current day Monte Carlo simulations. Since these are a consequence of the lattice discretisation and do not occur in the continuum version of the operator, we investigate in this paper to what extent the numerical evaluation of the overlap can be accelerated by making the Wilson-Dirac operator more continuum-like. Specifically, we study the effect of including the clover term in the Wilson-Dirac operator and smearing the link variables in the irrelevant terms. In doing so, we have obtained a factor of two speedup by moving from the Wilson action to a FLIC (Fat Link Irrelevant Clover) action as the overlap kernel.

\end{abstract}

\maketitle

\section{Introduction}

The overlap formalism \cite{overlap1, overlap2, overlap3, overlap4} leads, in the vector case, to a lattice formulation of QCD based on the overlap Dirac operator \cite{neuberger-massless}, given (in the massless case) by
\be
D_{\rm o}=\frac{1}{2a}\bigl(\,1+\gamma_5\epsilon(H_{\rm w})\bigr)
\qquad,\qquad\ \epsilon(H)=\frac{H}{\sqrt{H^2}},
\label{1}
\ee
($a$=lattice spacing) where
\be
H_{\rm w}=\gamma_5(D_{\rm w}-{\frac{m}{a}})
\label{2}
\ee
is a hermitian operator constructed from the Wilson-Dirac operator $D_{\rm w}$ \cite{wilson} with $m$ being a tuning parameter.\footnote{We are assuming that the Wilson parameter has been set to its canonical value $r=1$.} The free field propagator of $D_{\rm o}$ has the correct continuum limit and is free of doublers when $0<m<2$. Because of its origin in the overlap formalism, $D_{\rm o}$ has good chiral properties \cite{neuberger-overlap}; this can also be seen from the fact that it satisfies \cite{neuberger-moremassless} the Ginsparg-Wilson relation \cite{ginsparg-wilson}:
\be
\gamma_5D+D\gamma_5=2aD\gamma_5D.
\label{3}
\ee
Lattice Dirac operators satisfying this relation have an exact, lattice-deformed chiral symmetry \cite{luscher-chiral}, can have exact zero modes with definite chirality \cite{laliena}, as well as absence of mass renormalisation and other promising theoretical properties \cite{Hasenfratz(NPB),Shailesh,Giusti}.

The nice theoretical properties of the overlap Dirac operator come at a price: numerical evaluation of it via polynomial approximation is difficult due to the discontinuity at the origin of the matrix sign function $\epsilon(H)$. 
Practical methods have been developed in which $\epsilon(H)$ is approximated by a sum over poles $\epsilon_N(H)$, using either the so-called polar decomposition or the optimal rational polynomial approximation \cite{neuberger-practical, edwards-practical}, both of which take the form
\be
\epsilon_N(H) = H\left(c_0 + \sum_{k=1}^N \frac{c_k}{H^2+d_k}\right).
\label{4}
\ee
The two approximations only differ in their choice of coefficients $\{c_0,c_k,d_k\}$, and both are evaluated (indirectly) using a multi-shift Conjugate Gradient (CG) matrix inverter \cite{jegerlehner-multicg} to calculate their action on a vector. This is an iterative procedure where each iteration requires one evaluation of the matrix operator $H^2$ acting on a vector (i.e. two evaluations of $H$), and the number of iterations required to reach a given solution precision is proportional to the condition number of $H$, $\kappa(H) = |\lambda_{\rm max}/\lambda_{\rm min}|$, which is the ratio of the largest eigenvalue of $H$ to the smallest eigenvalue \cite{edwards-practical}.

Triangle inequalities lead to an upper bound \cite{neuberger-bounds} given by $|\lambda_{\rm max}|\le(8-m)/a$ for the operator $H_{\rm w}$ in Eq.\ (\ref{2}). The lower bound $|\lambda_{\rm min}|$ can be zero though. The lattice gauge fields for which $\lambda_{\rm min}=0$ form a subspace of measure zero in the space of all lattice gauge fields, so it is exceedingly unlikely that one would ever encounter them in a numerical simulation. However, our practical experience is that $|\lambda_{\rm max}| \lesssim 8$ while $|\lambda_{\rm min}|$ is often as small as $10^{-8}$. This results in an unacceptably large value for the condition number $\kappa(H)$. There is a way to get around this problem though \cite{edwards-chiral}. The typical spectrum of $H_{\rm w}$ is characterised by a handful of isolated low-lying eigenmodes, so one can project these out and deal with them explicitly. The condition number for the remaining part of the spectrum is then small enough that the approximation in Eq.\ (\ref{4}) becomes feasible. In practical simulations, after projecting out the isolated low-lying modes, $\epsilon_N(H_{\rm w})$ takes roughly speaking ${\cal O}(100-300)$ iterations to converge for $N\approx 14$, meaning that overlap fermions with the standard $H_{\rm w}$ are about ${\cal O}(200-600)$ times more expensive than standard Wilson fermions.

Obviously it is desirable to improve upon this situation in order to make simulations with overlap fermions more feasible. In this paper we investigate ways to do this by modifying the operator $H_{\rm w}$ in the overlap formula in Eq.\ (\ref{1}) so that its spectral properties are improved. The improvements we seek are twofold: (i) An upward shift in the magnitude of the low-lying eigenvalues of $H_{\rm w}$ so as to decrease the condition number, and (ii) a reduction in the {\em density} of low-lying eigenvalues, so as to make the projection method of Ref.\ \cite{edwards-chiral} more efficient. Furthermore, our aim is to produce an implementation of the overlap formalism that will perform efficiently on large-scale parallel computing architectures. On such architectures, the cost of internode communication is typically high compared to the cost of intranode computation. We therefore demand that our candidate $H$ be no less sparse than the Hermitian Wilson-Dirac operator, that is, possess at most nearest neighbour couplings. 

\section{Fermion actions}

The continuum version $H_{\rm c}=\gamma_5(\sd-\frac{m}{a})$ of $H_{\rm w}$ has the lower bound $|\lambda_{\rm min}|\ge\frac{m}{a}$ since $H_{\rm c}^2=-\sd^2+(\frac{m}{a})^2\ge(\frac{m}{a})^2$. Hence the near zero values of the lowest eigenvalues of $H_{\rm w}$ on equilibrium backgrounds at currently accessible $\beta$ are a result of the lattice discretisation. Our aim is to shift the lowest eigenvalues away from zero by making $H_{\rm w}\,$, or more specifically, the Wilson-Dirac operator $D_{\rm w}$ in $H_{\rm w}\,$, more continuum-like. In the framework of nonperturbative improvement of lattice operators (see, e.g., \cite{luscher-review}), $O(a)$ lattice artifacts in $D_{\rm w}$ are removed by adding the clover term of Ref.\ \cite{sheik-clover}. A simple heuristic argument for why this should should be beneficial in the present situation is the following. We write the Wilson-Dirac operator as 
\be
D_{\rm w}=\Sn+{\textstyle \frac{a}{2}}\Delta, \label{2.1}
\ee
where the naive lattice Dirac operator $\Sn$ and lattice Laplace operator $\Delta$ are given by
\begin{eqnarray}
a\nabla\four_{x,x'} \equiv a(\gamma^\mu\nabla_\mu)_{x,x'} &=& \frac{1}{2}\sum_\mu\big[ \gamma_\mu ( U_\mu(x)\delta_{x+e_\mu,x'} - U^\dagger_\mu(x-e_\mu)\delta_{x-e_\mu,x'} ) \big], \\
a^2\Delta_{x,x'} &=& 8\delta_{x,x'} - \sum_\mu\big[ U_\mu(x)\delta_{x+e_\mu,x'} + U^\dagger_\mu(x-e_\mu)\delta_{x-e_\mu,x'} \big].
\end{eqnarray}
The $\gamma$ matrices are chosen to be hermitian, so $\Sn$ is antihermitian and $\Delta$ is hermitian and positive. Define the operator $C$ by the relation
\be 
\Sn^2=\nabla^2+C\,.
\label{2.2}
\ee
(where $\nabla^2=\sum_{\mu}\nabla_{\mu}\nabla_{\mu}$). Then $C=\frac{1}{4}\lb\gamma_{\mu},\gamma_{\nu}\rb\lb\nabla_{\mu}, \nabla_{\nu}\rb$ is $\sim{\cal O}(a)$ and coincides with the usual clover term (with coefficient $c_{\rm sw}=1$, the tree level value) up to an ${\cal O}(a^2)$ term. Here and in the following ${\cal O}(a^p)$ denotes a lattice operator whose leading term in a formal expansion in powers of the lattice spacing is $\sim{}a^p$. Now, setting the parameter $m$ in $H_{\rm w}$ to its canonical value $m=1$ we have
\be
H_{\rm w}^2=(D_{\rm w}-{\textstyle \frac{1}{a}})^*(D_{\rm w}-{\textstyle \frac{1}{a}})
=-\Sn^2-\Delta+({\textstyle \frac{a}{2}}\Delta)^2
+{\textstyle \frac{a}{2}}\lb\Delta,\Sn\rb+{\textstyle \frac{1}{a^2}}. \label{2.3}
\ee
Straightforward calculations show that $\Delta+\nabla^2\sim{\cal O}(a^2)$ and $\lb\Delta,\Sn\rb\sim{\cal O}(a)\,$; hence, by Eq.\ (\ref{2.2}), we have $\;\Sn^2+\Delta=C+{\cal O}(a^2).$ Hence we obtain the lower bound
\be
H_{\rm w}^2\;\ge\;{\textstyle \frac{1}{a^2}}-C-{\cal O}(a^2)
={\textstyle \frac{1}{a^2}}-{\cal O}(a).
\label{2.4}
\ee
Thus the lower bound $\frac{1}{a^2}$ on the continuum version of $H_{\rm w}^2$ is spoiled in the lattice case by an ${\cal O}(a)$ term. If we now add $C$ to $\Delta$ in (\ref{2.1}), i.e. replace
\be
D_{\rm w}\;\rightarrow\;D_{\rm cw}\equiv\Sn+{\textstyle \frac{a}{2}}(\Delta+C)
\label{2.5}
\ee
we find
\be
H_{\rm cw}^2&=&(D_{\rm cw}-{\textstyle \frac{1}{a}})^*(D_{\rm cw}-{\textstyle \frac{1}{a}})
\nonumber \\
&=&-\Sn^2-(\Delta+C)+({\textstyle \frac{a}{2}}(\Delta+C))^2
+{\textstyle \frac{a}{2}}\lb\Delta+C\,,\Sn\rb+{\textstyle \frac{1}{a^2}}
\nonumber \\
&\ge&{\textstyle \frac{1}{a^2}}-{\cal O}(a^2).
\label{2.6}
\ee
Hence the ${\cal O}(a)$ term $(-C)$ in Eq.\ (\ref{2.4}) has dropped out and the continuum lower bound $\frac{1}{a^2}$ is now only spoiled by an ${\cal O}(a^2)$ term. 

However, it is well-known that adding a clover term only improves the chiral properties of the Wilson-Dirac operator on smooth backgrounds, and that the localisation of the real eigenvalues of $D_{\rm cw}$ is actually {\em worse} than for $D_{\rm w}$ on typical gauge backgrounds generated in Monte Carlo simulations \cite{gattringer-clover, DeGrand, stephenson}. This suggests that to further improve the chiral properties of $D_{\rm cw}$ we should consider smoothing the lattice gauge field. In Ref.\ \cite{DeGrand}, DeGrand {\em et. al.} found that a significant improvement in the chiral properties can be achieved by applying an APE smearing procedure \cite{ape-one,ape-two, derek-apesmearing, derek-smooth} to the link variables, leading to a fat link version of $D_{\rm cw}$. (The idea of using fat links in fermion actions was first explored by the MIT group \cite{MIT}.)
More recently, Zanotti {\em et.al.} have shown in \cite{zanotti-hadron} that such improvement can be achieved by smearing only the link variables appearing in the irrelevant operators, i.e. in the Wilson and clover terms. This has the advantage of preserving the short distance quark interactions. (The idea of using fat links in the irrelevant operators had been independently suggested previously in Ref.\ \cite{neuberger-bounds}.)

Motivated by the preceding discussion, we compare the evaluation of the usual overlap Dirac operator with the operators obtained by replacing $H_{\rm w}$ in the overlap formula of Eq.\ (\ref{1}) with the following variants. (The lattice spacing has been set to $1$ unless specified otherwise):

\noindent (i) {\em Hermitian Wilson-Dirac operator with clover term}:
\be
H_{\rm cw}(m,c_{\rm sw})=\gamma_5\big( \nabla\four + 
\frac{1}{2}(\Delta - \frac{c_{\rm sw}}{2}\sigma\cdot F) - m \big). 
\label{2.7}
\ee
where
\be
\sigma_{\mu\nu}&=&\frac{1}{2}[\gamma_\mu,\gamma_\nu], \quad F_{\mu\nu}(x) =\frac{1}{2}\big(C_{\mu\nu}(x)-C_{\mu\nu}^\dagger(x)\big), \\
C_{\mu\nu}(x)&=& \frac{1}{4}\big( U_{\mu\nu}(x) + U_{-\nu\mu}(x) + U_{\nu-\mu}(x) + U_{-\mu-\nu}(x) \big).
\ee

\noindent (ii) {\em Fat link Hermitian Wilson-Dirac operator, both with and without clover term}:
\begin{eqnarray}
H_{\rm fw}(m,\alpha n_{\rm ape}) &=& \gamma_5\big( \nabla\four + \frac{1}{2}\Delta^{(\alpha n_{\rm ape})} - m \big), \label{2.8} \\
H_{\rm fcw}(m,c_{\rm sw}, \alpha n_{\rm ape}) &=& \gamma_5\big( \nabla\four + \frac{1}{2}(\Delta^{(\alpha n_{\rm ape})} - \frac{c_{\rm sw}}{2}\sigma\cdot F^{(\alpha n_{\rm ape})}) - m \big). \label{2.9}
\end{eqnarray}
where APE-smearing is carried out on the individual links in the irrelevant
operators by making the replacement
\eqn{
\label{eqn:apesweep} U_\mu(x) \rightarrow U^{(\alpha)}_\mu(x) = {\mathcal P}\left( (\alpha -1) U_\mu(x) + \frac{\alpha}{6}\sum_{\pm\nu\neq\mu} U_\nu(x)U_\mu(x+ae_\nu)U^\dagger_\nu(x+ae_\mu) \right).
}
Here ${\mathcal P}$ denotes projection of the RHS of Eq.\ (\ref{eqn:apesweep}) back to the SU(3) gauge group. That is, each link is modified by replacing it with a combination of itself and the surrounding staples to give a set of ``fat links''.  The means by which one projects back to SU(3) is not unique. We choose an SU(3) matrix $U^{(\alpha)}_\mu(x)$ such that the gauge invariant measure ${\rm Re}{\rm Tr}(U^{(\alpha)}_\mu(x)X^\dagger_\mu(x))$ is maximal, where $X_\mu(x)$ is the smeared link before projection, that is $U^{(\alpha)}_\mu(x)\equiv {\mathcal P}X_\mu(x)$. As the process of APE-smearing removes short-distance physics, it is preferable to only smear the irrelevant operators. Throughout this work ``fat'' means APE smearing of links in irrelevant terms only. 
Here $\alpha$ is the smearing fraction and $n_{\rm ape}$ is the number of smearing sweeps (\ref{eqn:apesweep}) we perform. As shown in \cite{derek-smooth}, we can effectively reduce the two-dimensional parameter space $(\alpha,n_{\rm ape})$ to a one-dimensional space that depends soley on the product $\alpha n_{\rm ape}\,$, and this is reflected in the notation in Eqs. (\ref{2.8})--(\ref{2.9}).

Finally, as in \cite{zanotti-hadron}, we can perform tadpole or mean-field improvement (MFI) \cite{lepage-mfi} to bring our links closer to unity. This consists of updating each link with a division by the mean link, which is the fourth root of the average plaquette,

\eqn{ u_0 = \langle {\rm \frac{1}{3}ReTr } U_{\mu\nu}(x) \rangle_{x,\mu<\nu}^{\frac{1}{4}}. }
 
In the case of $H_{\rm w}$ and $H_{\rm fw}$, mean-field improvement has little effect, entering in only as a single power in both cases. For $H_{\rm w}$, mean field improvement effectively changes the value of $m$ and renormalises the Wilson parameter $r$. In the case of $H_{\rm fw}$ it has a similar effect but we have two mean link values, one for the untouched set of links and one for the smeared set. However, $u_0$ enters in as the fourth power in front of the clover term, effectively raising $c_{\rm sw}$ towards its non-perturbative value. 
Hence our final two variants of $H_{\rm w}$ are the following.

\noindent (iii) {\em MFI clover Hermitian Wilson-Dirac operator, both with and without fat links}:
\begin{eqnarray}
H^{\rm mfi}_{\rm cl}(m,c_{\rm sw}) &=& \gamma_5\big( \frac{1}{u_0}\nabla\four + \frac{1}{2}(\frac{1}{u_0}\Delta - \frac{c_{\rm sw}}{2u^4_0}\sigma\cdot F) - m \big), \\
H^{\rm mfi}_{\rm fcl}(m,c_{\rm sw}, \alpha n_{\rm ape}) &=& \gamma_5\big( \frac{1}{u_0}\nabla\four + \frac{1}{2}(\frac{1}{u^{\rm fl}_0}\Delta^{(\alpha n_{\rm ape})} - \frac{c_{\rm sw}}{2(u^{\rm fl}_0)^4}\sigma\cdot F^{(\alpha n_{\rm ape})}) - m \big),
\end{eqnarray}
where we have differentiated the mean link $u_0$ for the untouched links and $u_0^{\rm fl}$ for the fat links.  We refer to the MFI fat clover action as the FLIC (Fat-Link Irrelevant Clover) action. The FLIC action was recently introduced and studied in Ref \cite{zanotti-hadron}. If followed by a number (e.g. FLIC12) this denotes the number of APE-smearing sweeps (at $\alpha = 0.7$) used in the action.

Before proceeding to the numerical results it is worth pointing out that the previous analytical results on the locality \cite{luscher-locality} and continuum limit of the axial anomaly \cite{Fujikawa, Suzuki, Adams-axial, Kikukawa} and index \cite{adams} of the overlap Dirac operator continue to hold when $H_{\rm w}$ is replaced by any of the variants given above in the overlap formula. In the case of the axial anomaly and index, this is essentially because the leading order term in the expansion of commutators of the covariant finite difference operators in powers of the lattice spacing is unchanged, and the variants of $H_{\rm w}$ all coincide with $H_{\rm w}$ in the free field case. Regarding locality, the admissibility bounds of \cite{neuberger-bounds,luscher-locality} on the plaquette variables get modified somewhat when the different variants of $H_{\rm w}$ are used. In light of the heuristic arguments above and our numerical results below, we expect that it should be posible to derive improved, (i.e., less restrictive), bounds in these cases, although so far we have not been able to show this.

We also mention that more general variants of the overlap Dirac operator have been considered where one starts with an approximate solution to the Ginsparg-Wilson relation and then gets an exact solution by substituting into the overlap formula \cite{Bietenholz-fast,Bietenholz-GW,Hasenfratz-NPPS,Hasenfratz-IJMP,DeGrand(overlap)}.\footnote{Specifically, if $D_{\rm approx}$ is some approximate solution to the GW relation then $A=1-D_{\rm approx}$ satisfies $A^*A\approx 1$. An exact solution $D$ to the GW relation, which is approximately equal to $D_{\rm approx}\,$, is then obtained via the overlap formula by setting $D=1-\frac{A}{\sqrt{A^*A}}$.} This has led to variants of the overlap action which are both easier to evaluate and more local than the original. However, it is not clear whether such general operators will have the good topological properties of the standard overlap Dirac operator, namely exact zero-modes with definite chirality in topologically nontrivial backgrounds, (c.f. the counter-example of Chiu \cite{chiu-axial, chiu-index}). This is important in connection with the lattice implementation of the Witten-Veneziano formula for the $\eta'$ mass with GW fermions \cite{Giusti}, since for the argument to work it is essential that the would-be zero modes are exact zero modes.

\section{Spectral Flow Comparison}

In order to test the merits of each of our proposed actions, we first calculate the spectral flow of each of them to see if our reasoning regarding their low-lying spectra is valid. From the quadratic form of the lower bounds as a function of $m$, and based upon results given in Ref.\ \cite{edwards-spectral}, we expect there to be some peak value of $m$ for which the gap around zero is the largest. We calculated the flow of the lowest 15 eigenvalues as a function of $m$ for an ensemble of 10 mean-field improved Symanzik configurations at $\beta = 4.38$ and size $8^3\times16$. The following flow graphs allow us to see the $m$ value for the biggest gap, and also allow us to compare the different actions. As we are interested in the magnitude of the low-lying values rather than their sign, we plot $|\lambda|$ vs $m$.

\begin{figure}[!htb]
\includegraphics[height=0.48\textwidth, width=0.28\textheight, angle=90 ]{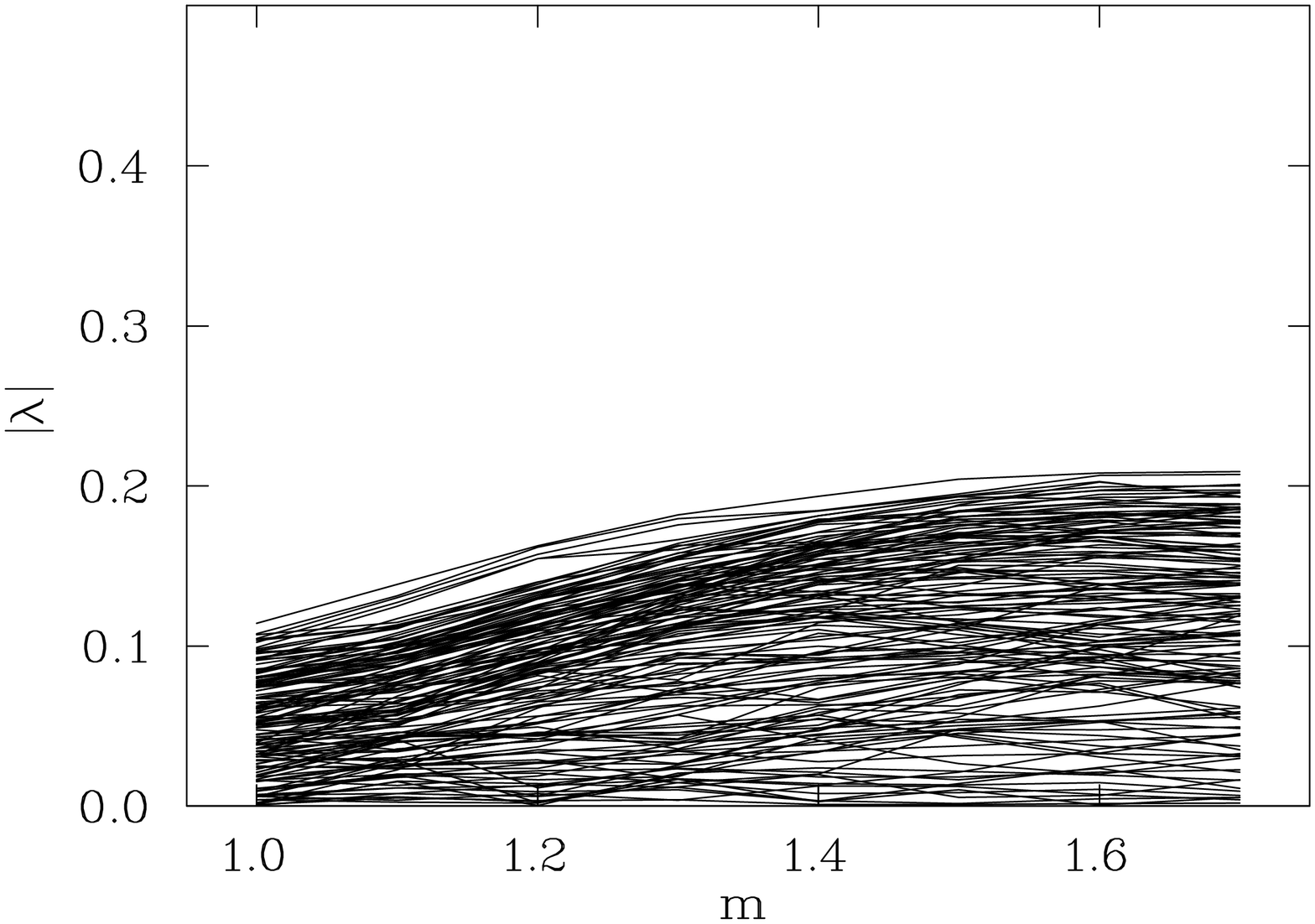}
\includegraphics[height=0.48\textwidth, width=0.28\textheight, angle=90 ]{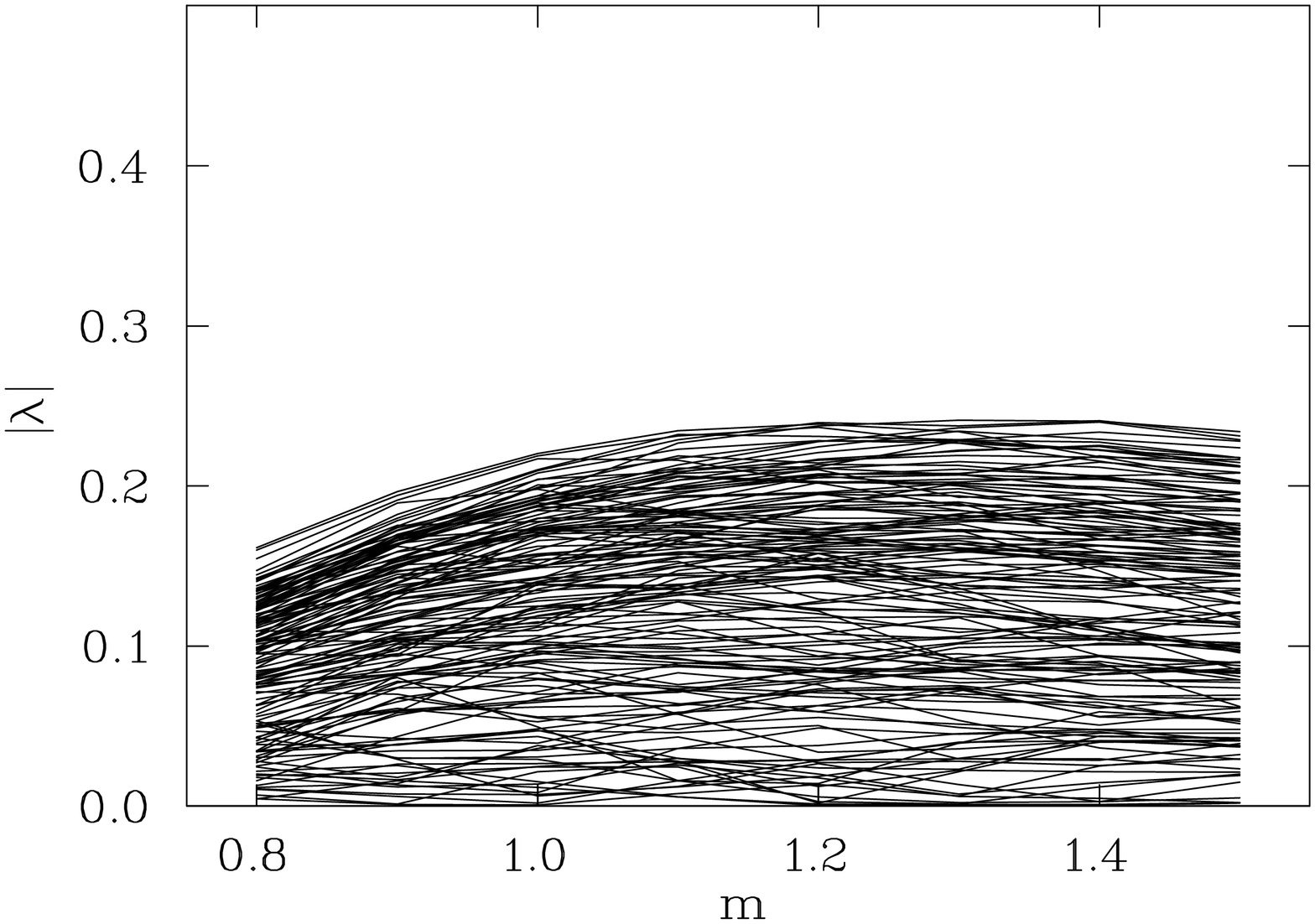}
\caption{\label{fig:wilson} Spectral flow of the Wilson action (left) and the clover action (right) at $\beta=4.38$.}
\end{figure}

We begin by examining the flow of the Wilson and clover action in Figure \ref{fig:wilson}. We see the Wilson spectrum is very poor, with a high density of very small eigenmodes and no gap away from zero. The addition of the clover term (at $c_{\rm sw}=1$) provides some improvement, shifting the flow upwards and moving the peak values towards $m=1$ as expected. The presence of many small eigenmodes persists however, although their density is clearly reduced.

\begin{figure}[!thb]
\includegraphics[height=0.48\textwidth, width=0.28\textheight, angle=90 ]{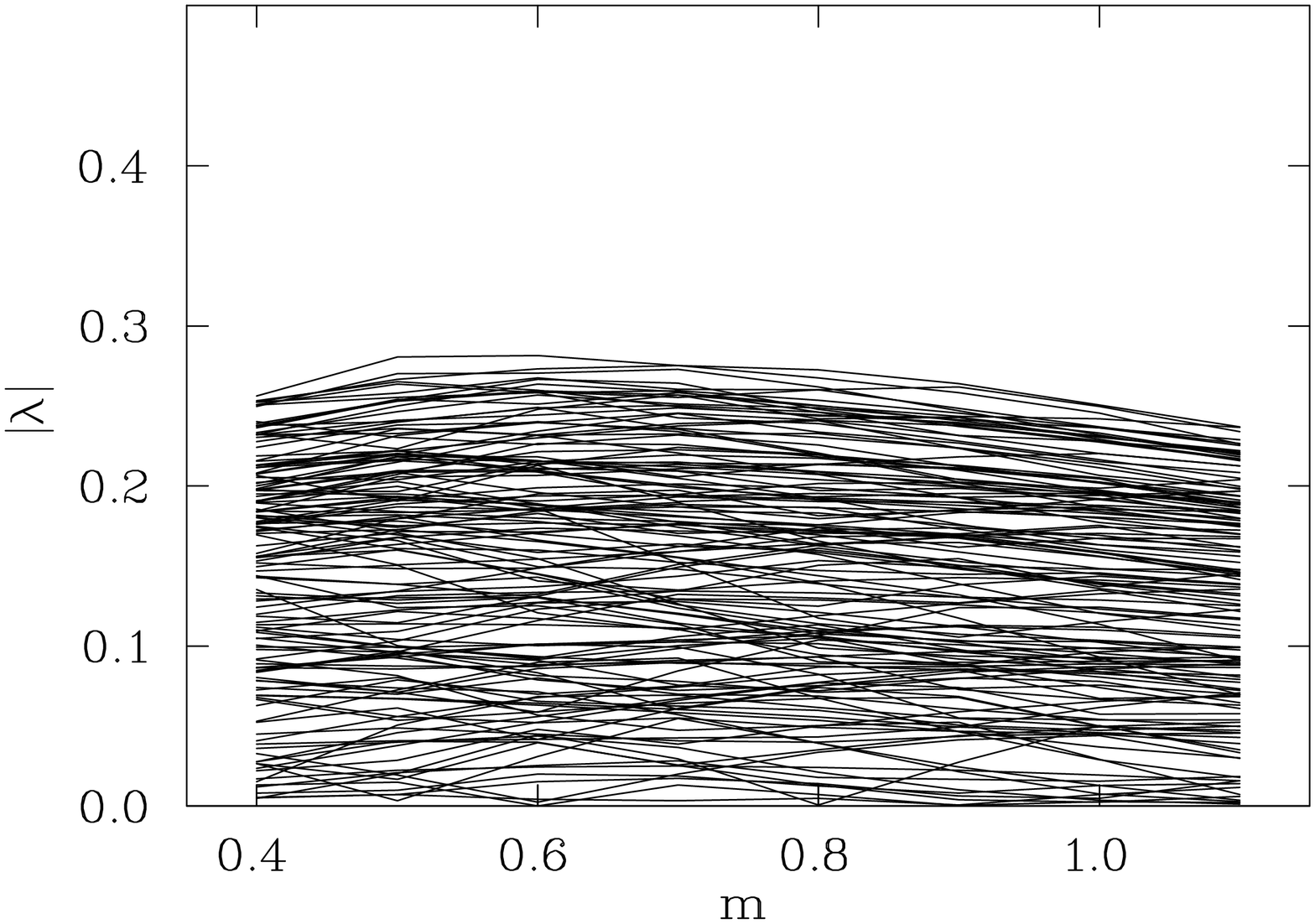}
\includegraphics[height=0.48\textwidth, width=0.28\textheight, angle=90 ]{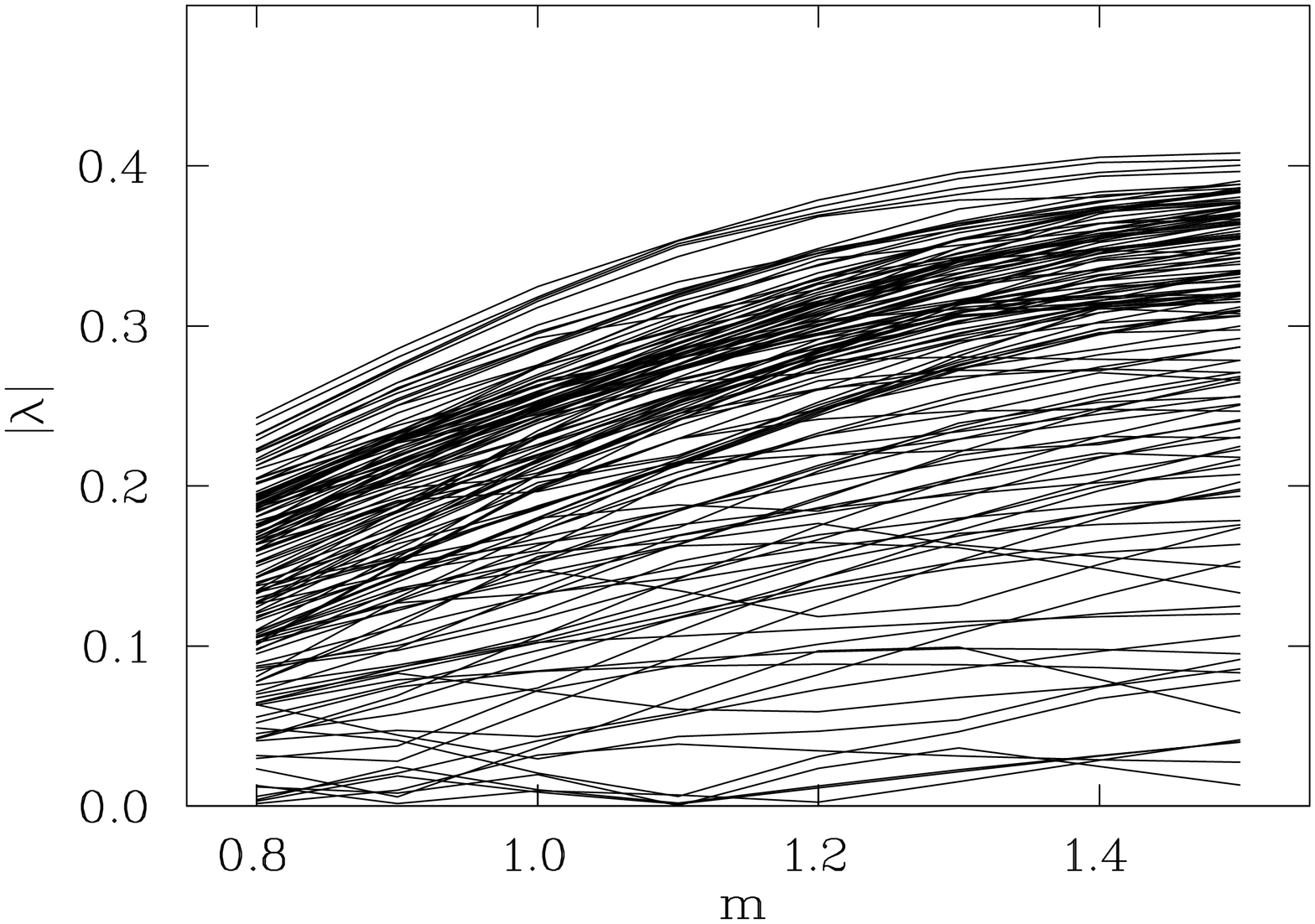}
\caption{\label{fig:mficlover} Spectral flow of the MFI clover action (left) and the fat Wilson action (right) at $\beta=4.38$. }
\end{figure}

In Figure \ref{fig:mficlover} we examine the MFI clover and fat Wilson actions. Mean field improvement assists the basic clover action somewhat, spreading the spectrum upwards, although the lowest modes are not raised significantly. The mass value at which the low-lying density is minimised has moved significantly away from $m=1.2$ to around $m=0.6$. As mentioned earlier, essentially all MFI does in this case is to change the value of $c_{\rm sw}$ to $1.0/u^3_0$, pushing it towards its non-perturbative value. Modifying the Wilson action by smearing the irrelevant operators provides a considerable improvement. While there are still some small modes present, their density has been greatly reduced, and the spectral flow now has a clear division between the isolated low-lying modes and the modes where the spectral density becomes high which are well separated from zero. Smearing was performed with $\alpha=0.7$ and $ n_{\rm ape}=12$ smearing sweeps.
   
\begin{figure}[!thb]
\includegraphics[height=0.48\textwidth, width=0.28\textheight, angle=90 ]{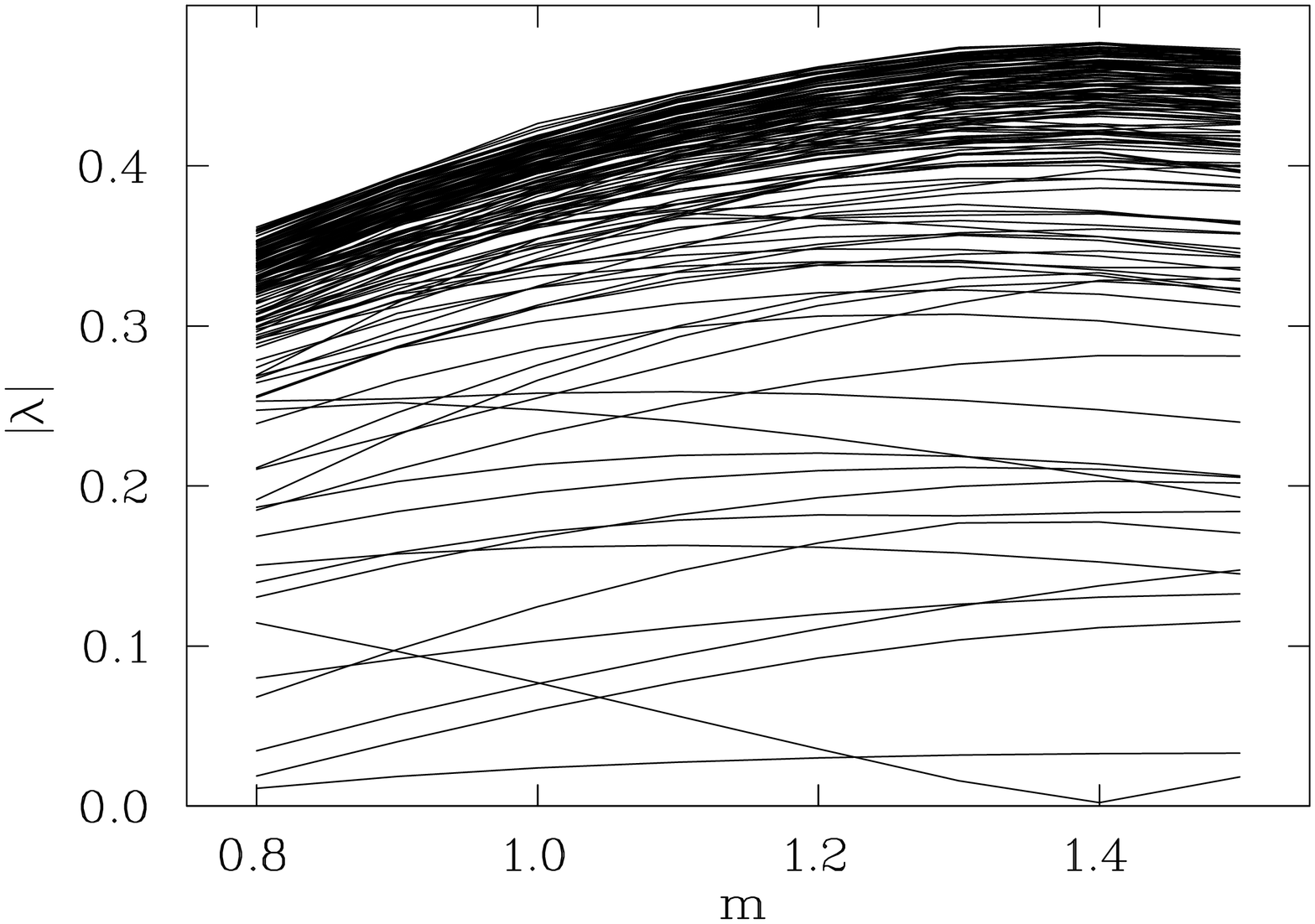}
\includegraphics[height=0.48\textwidth, width=0.28\textheight, angle=90 ]{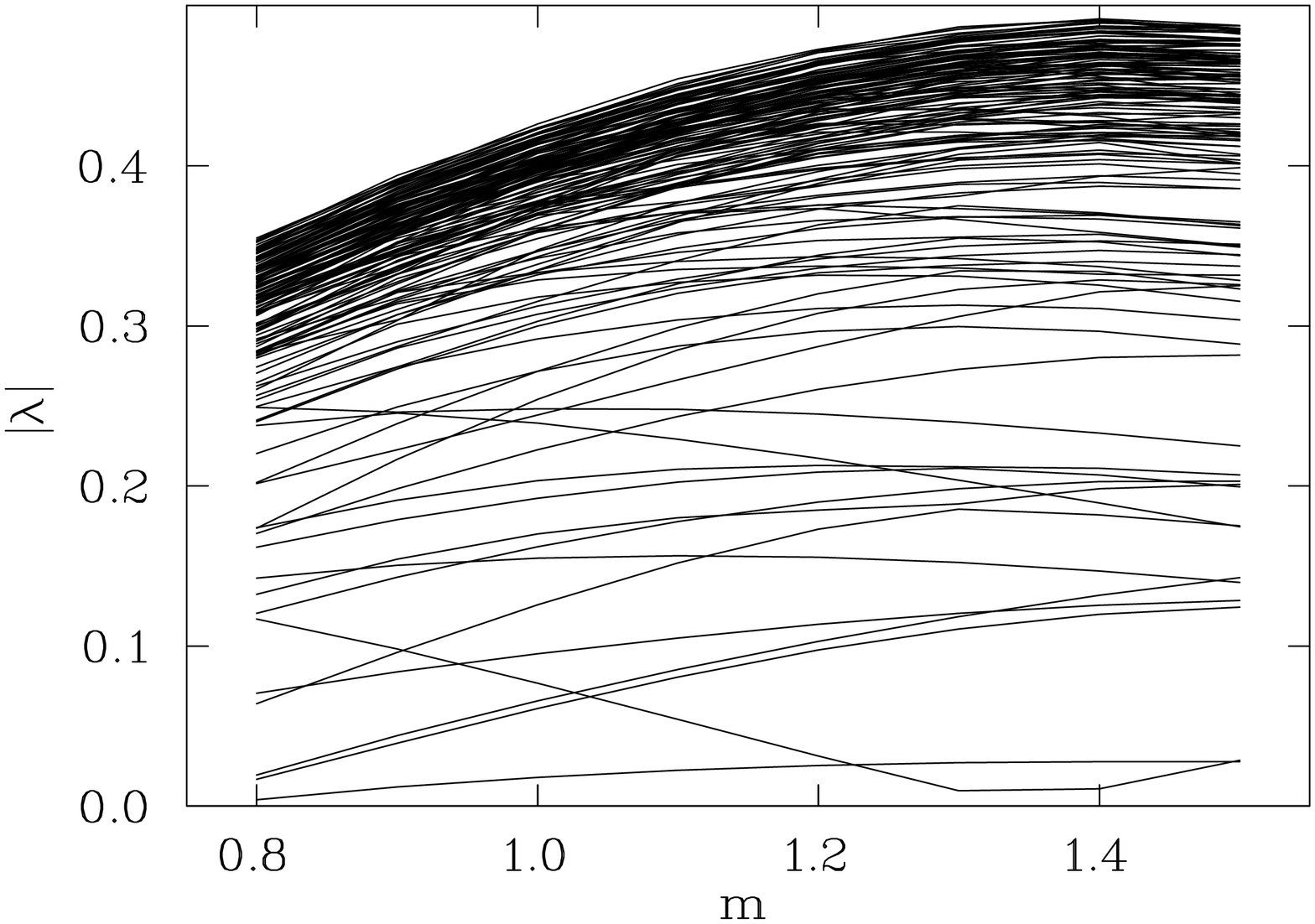}
\caption{\label{fig:flic} Spectral flow of the fat clover action (left) and FLIC12 action (right) at $\beta=4.38$. }
\end{figure}

Results for the fat clover and FLIC12 actions are shown in Figure \ref{fig:flic}. The spectral flow of the fat clover action clearly demonstrates the superiority of clover-improved actions. The gap around zero is enhanced again over the fat Wilson action, and the number of isolated low-lying modes is significantly reduced. As the fat links are already close to unity, the addition of mean field improvement only affects the fat clover flow slightly, raising the gap around zero a little and spreading the eigenvalues upwards slightly also. The low-lying density is again very good in this case and far superior to that of the Wilson action.

To confirm our results we choose the Wilson action as a ``baseline'' and compare it against the FLIC action (the best of the alternative actions) on a larger, finer lattice, $12^3\times24$ at $\beta=4.60$. This time we only use 4 smearing sweeps in the FLIC action since FLIC4 has less fattening and is the choice used in actual simulations\cite{zanotti-hadron}. We see that the Wilson action benefits significantly from the smaller lattice spacing, as there is now a visible separation from zero before the modes become dense. The FLIC action has the same characteristics as on the coarser lattice, but it now has a peak separation of the dense modes from zero of around 0.45!

\begin{figure}[!thb]
\includegraphics[height=0.48\textwidth, width=0.28\textheight, angle=90 ]{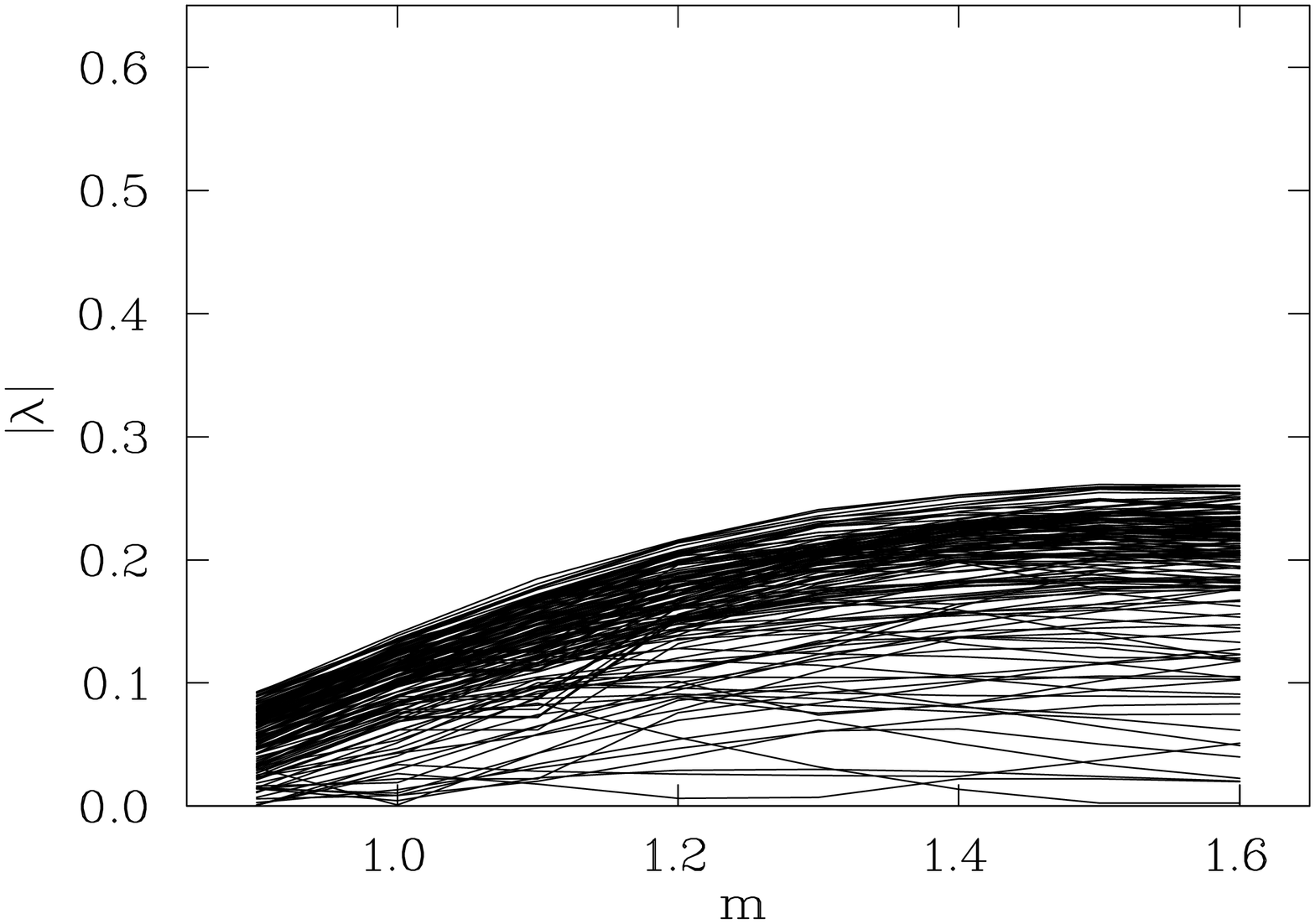}
\includegraphics[height=0.48\textwidth, width=0.28\textheight, angle=90 ]{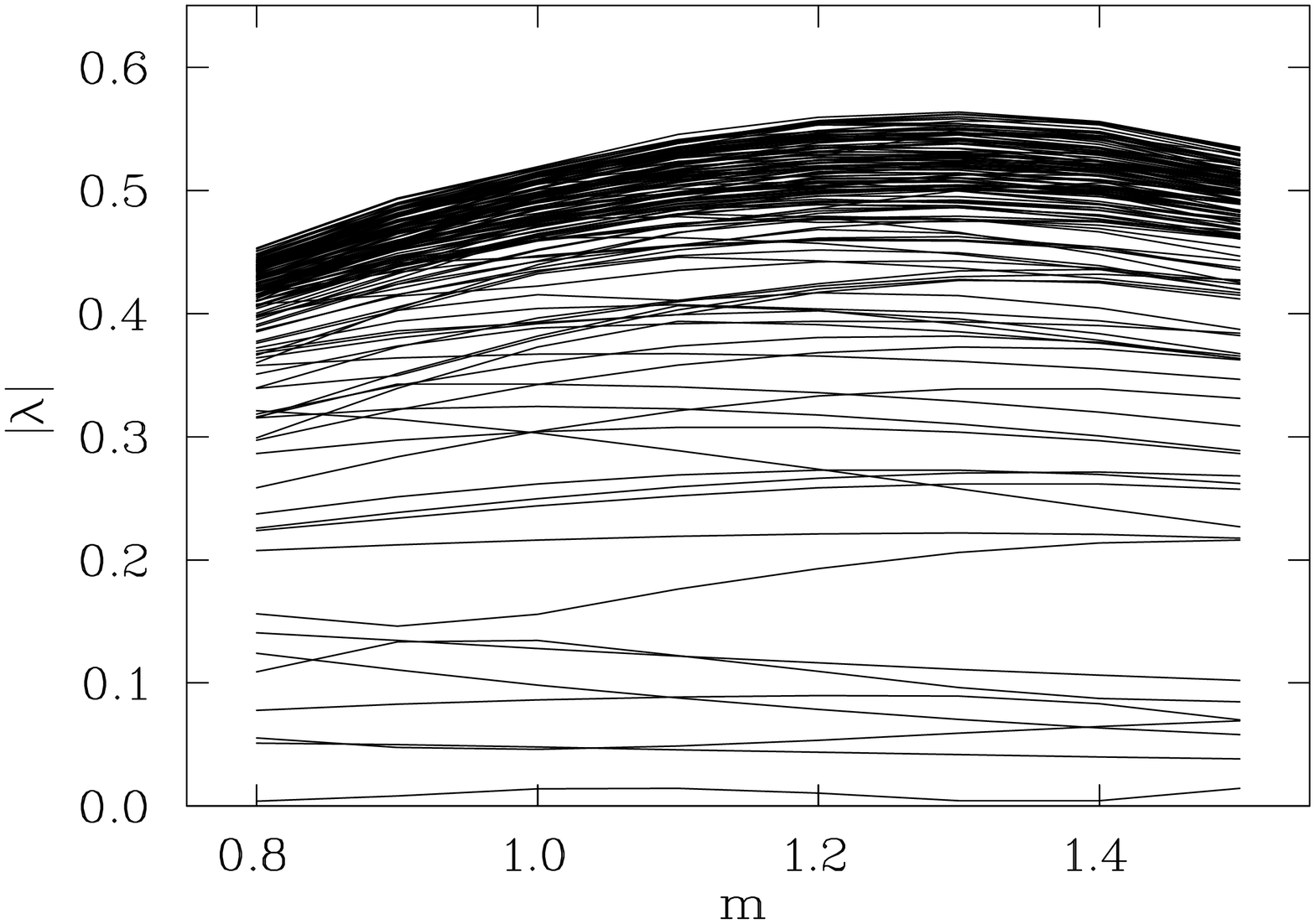}
\caption{Spectral flow of the Wilson action (left) and FLIC4 action (right) at $\beta=4.60$. }
\end{figure}

\begin{figure}[!thb]
\includegraphics[height=0.48\textwidth, width=0.28\textheight, angle=90 ]{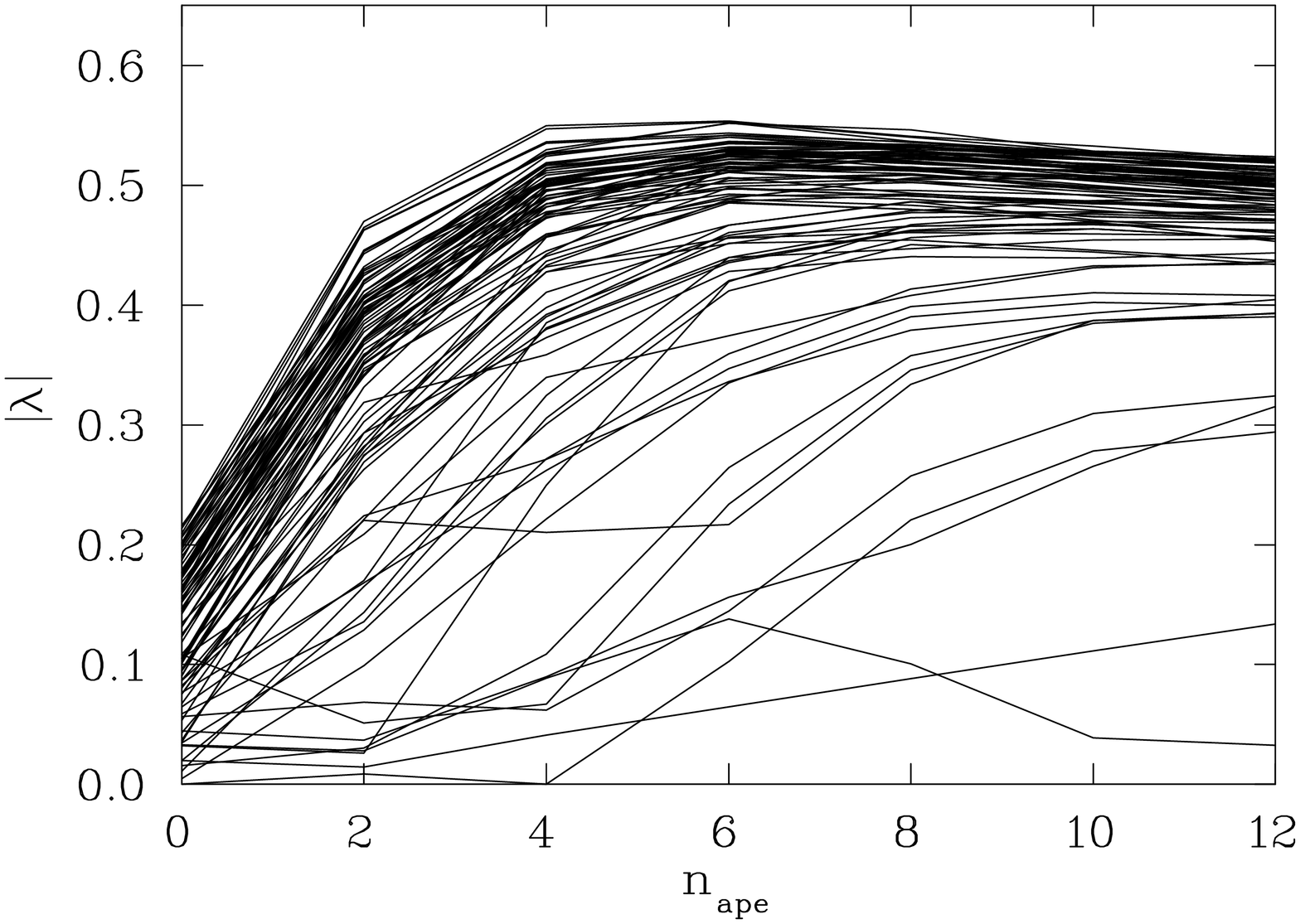}
\includegraphics[height=0.48\textwidth, width=0.28\textheight, angle=90 ]{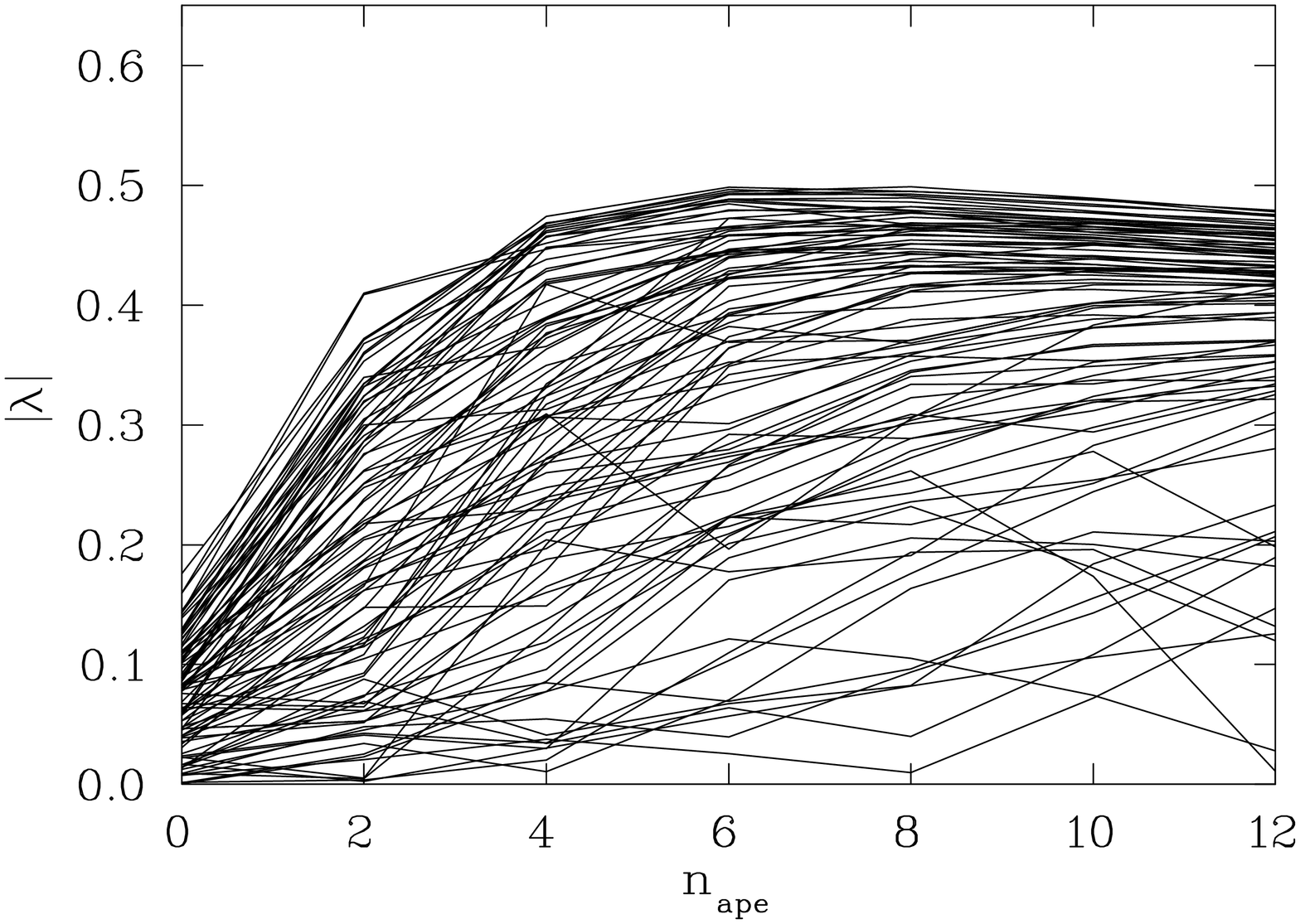}
\caption{Dependence of the FLIC spectrum at $\beta=4.60, m=1.35$ (left) and $\beta=4.38,m=1.45$ (right) on the number of APE smearing sweeps.}
\end{figure}

Additionally, we tested the dependence of the FLIC action upon the amount of smearing done. As stated in \cite{derek-smooth}, we only effectively need to vary the product $\alpha n_{\rm ape}$, so we fix $\alpha$ at 0.7 and vary $n_{\rm ape}$ between 0 and 12. We observe that the initial 4-6 sweeps have a significant effect, but past 6 sweeps the effect is marginal, with the low lying density remaining roughly constant and the eigenvalues being compressed very slightly downwards.
     
\section{Results}

Having obtained some understanding of the low-lying spectra of the various actions via the flow diagrams, we now turn to quantitative comparisons. Firstly we examine the condition number, $\kappa$, of the different actions as a function of $m$. We show below the condition numbers having projected out the lowest 5 eigenmodes and the lowest 15 eigenmodes on the 2 lattices that we used. The points are the mean condition numbers across the ensembles, and the error bars indicate the minimum and maximum condition numbers, giving an idea of the variation in $\kappa$. The smeared irrelevant-term actions here used 12 APE sweeps at $\alpha=0.7$ for the coarse lattice and 4 sweeps for the fine lattice. Some points are offset horizontally for clarity.

\begin{figure}[!hbt]
\includegraphics[height=0.48\textwidth, width=0.31\textheight, angle=90 ]{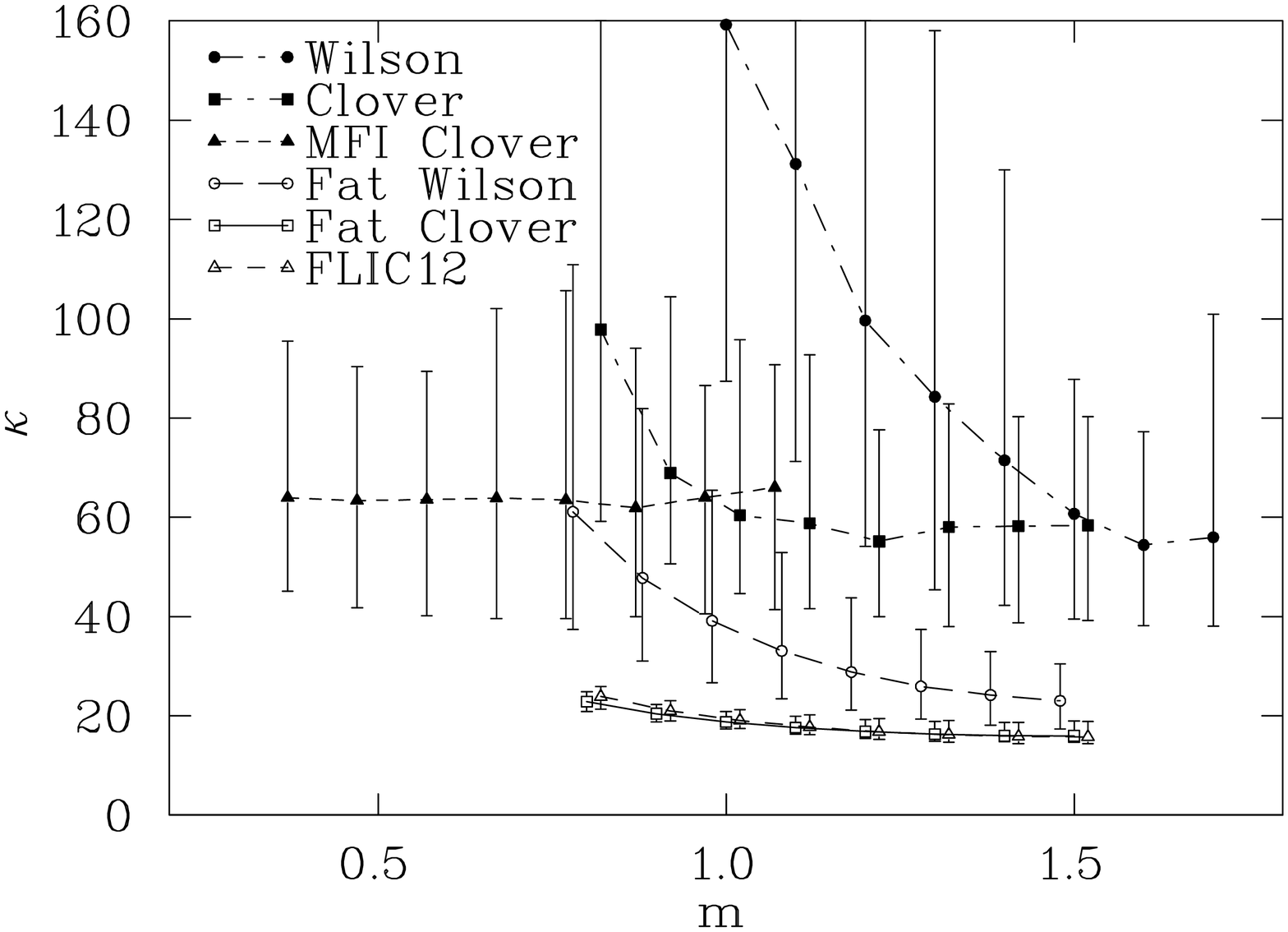}
\includegraphics[height=0.48\textwidth, width=0.31\textheight, angle=90 ]{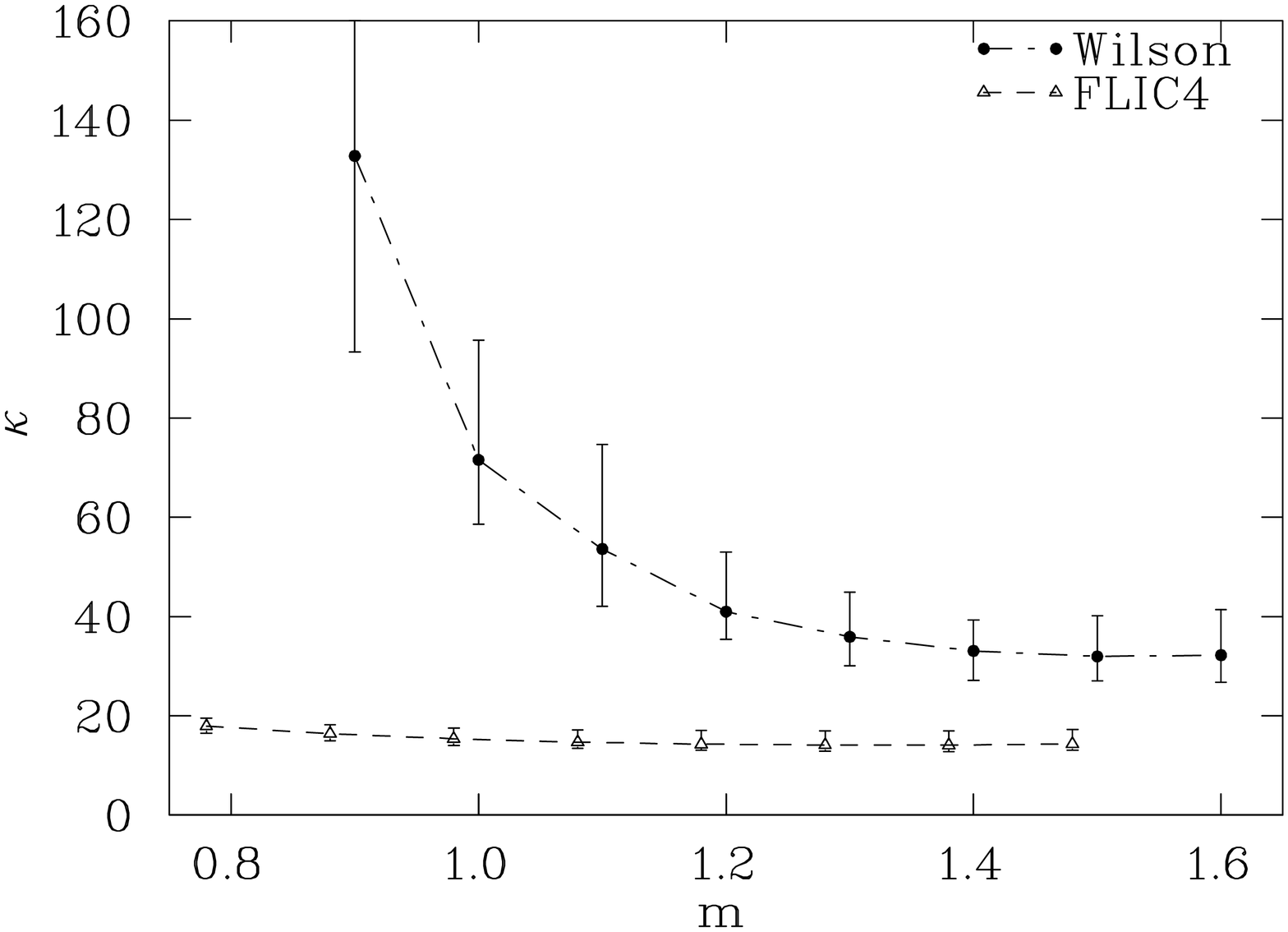}

\includegraphics[height=0.48\textwidth, width=0.31\textheight, angle=90 ]{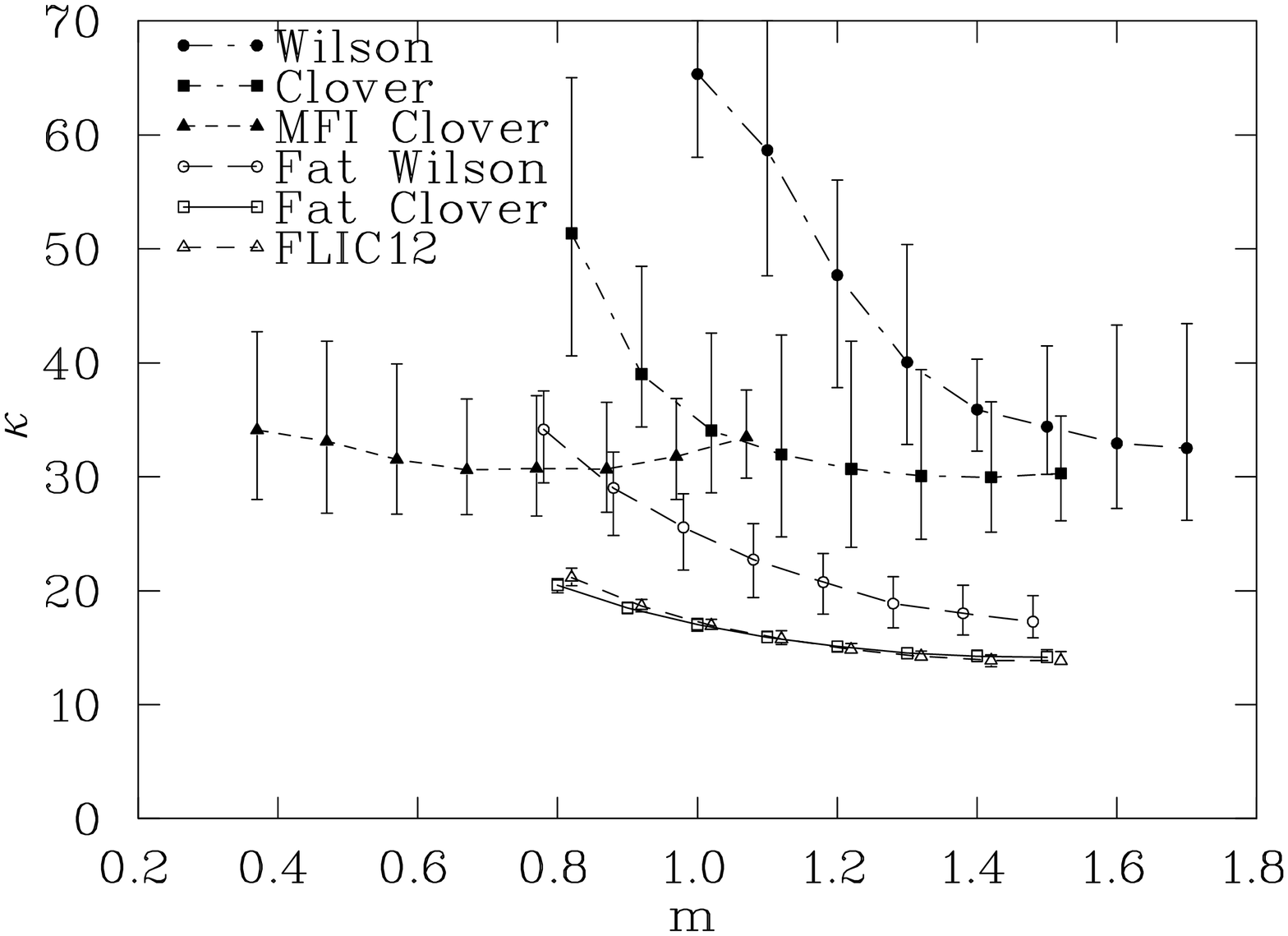}
\includegraphics[height=0.48\textwidth, width=0.31\textheight, angle=90 ]{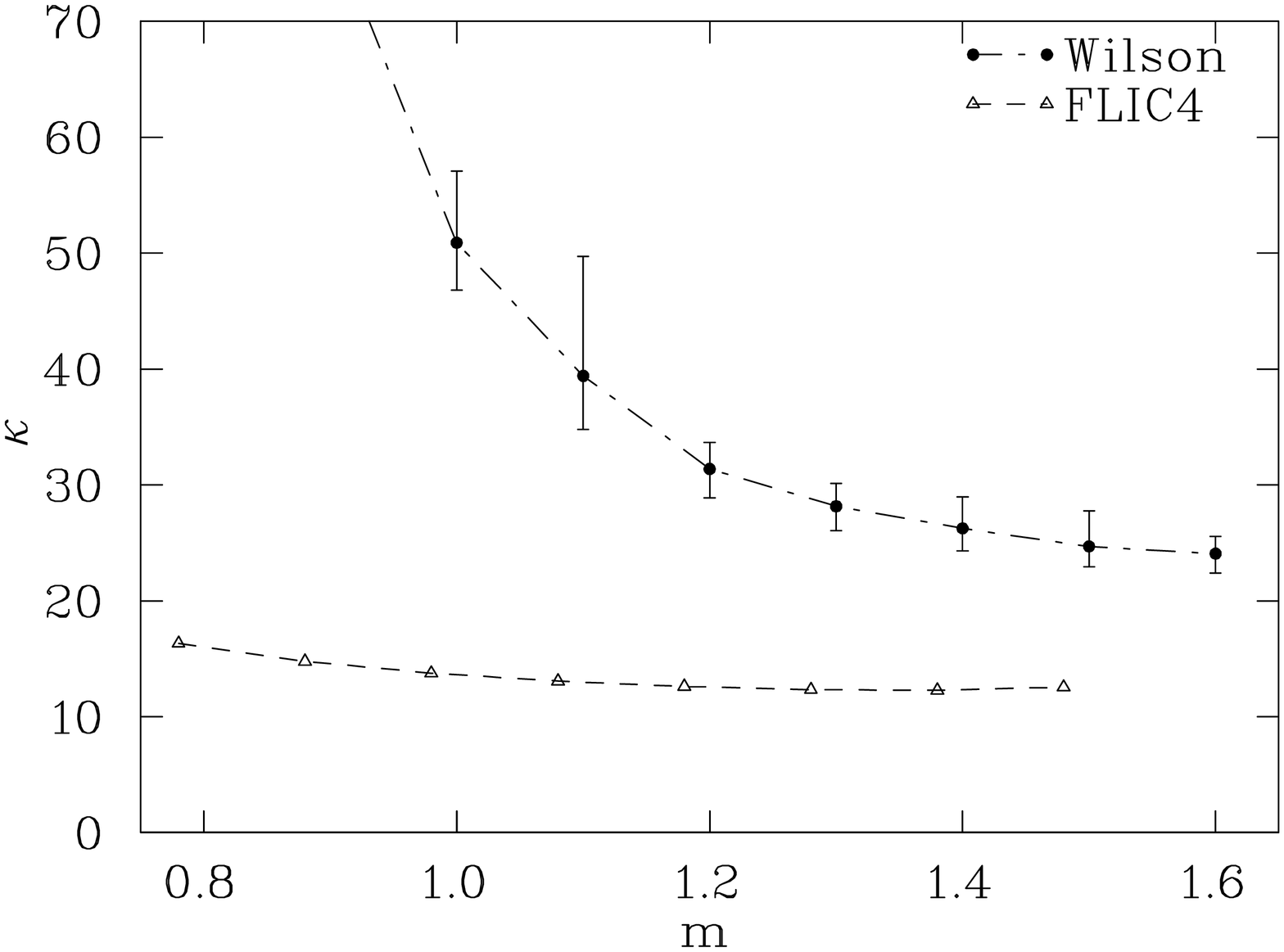}

\caption{Condition numbers of the various actions. (Top-left) Results for $\beta=4.38$ with 5 projected modes. (Bottom-left) Results for $\beta=4.38$ with 15 projected modes. (Top-right) Results for $\beta=4.60$ with 5 projected modes. (Bottom-right) Results for $\beta=4.60$ with 15 projected modes.}
\end{figure}

Two things are immediately noticeable. Firstly, the smeared irrelevant-term actions are much better conditioned than the unsmeared actions, and secondly, the variation of $\kappa$ between configurations is less. It should be noted that the variation (error bars) are displayed for all actions, but are smaller than the plot symbol at some points of the fat clover and FLIC lines. Projecting out an additional 10 eigenvalues has a significant effect on the unsmeared actions, but relatively little effect on the smeared actions due to reduction in the number of isolated low-lying values. In terms of condition number, the fat clover and FLIC actions are clearly and significantly superior to the other actions, with the FLIC action possessing a (slight) edge over the fat clover which arises from the mean field improvement.

As the clover term is quite fast to evaluate, we discard the fat Wilson as a candidate action at this point as it is the least well-conditioned of the smeared actions. Given the similarity between the clover-improved actions with and without mean-field improvement, we focus on the MFI clover and FLIC actions. We now compare in detail the performance for three actions: the Wilson, MFI clover and FLIC. To see how improving the condition number translates into a saving in CG iterations, we calculated the number of Multi-CG iterations required to evaluate $D_{\rm o}$ once across the ensemble for each of these actions, using some typical simulation parameters. 

\begin{table}[!htb]
\begin{center}
\begin{tabular}{lccccc}
Action & $\beta$ & Projections & Mean & Min & Max \\
\hline
Wilson & 4.38 & 15 & 219 & 188 & 253 \\
 & 4.60 & 15 & 202 & 190 & 212 \\
MFI clover & 4.38 & 15 & 200 & 178 & 240 \\
FLIC12 & 4.38 & 10 & 92 & 89 & 100 \\
FLIC6 & 4.38 & 10 & 90 & 86 & 101 \\
FLIC4 & 4.60 & 15 & 109 & 106 & 112 \\
\end{tabular}
\end{center}
\caption{\label{tab:convergence} Conjugate Gradient (CG) Iterations needed for a single evaluation of $\epsilon_N(x)$ using actual simulation parameters.}
\end{table}

The Wilson and MFI clover are tested using the 14th order optimal rational polynomial (ORP) approximation \cite{edwards-chiral}. The improved condition number of the FLIC actions allows us to use the 12th order polar decomposition, chosen to give a maximum deviation from $\epsilon(x)$ of less than $10^{-6}$ compared to the $3.1\times 10^{-5}$ of the 14th order ORP. The N$^{th}$
 order polar decomposition is specified by
\eqn{ c_0 = 0,\quad c_k = \frac{1}{N \cos^2(\frac{\pi}{2N}(k-\frac{1}{2}))}, \quad d_k = \tan^2(\frac{\pi}{2N}(k-\frac{1}{2})).}
 Low-lying modes are projected out where necessary. The sign function solution is calculated to a precision of $10^{-6}$ across the fine ensemble and the coarse ensemble used above. The value of $m$ is chosen differently for each of the actions to optimise $\kappa$.
 Given the relative lack of improvement in using the MFI clover action compared to the Wilson, we discard it at this point and concentrate on comparing the Wilson and FLIC actions. As the results in Table \ref{tab:convergence} show, the FLIC action is by far the best in terms of convergence with a reduction in iterations compared to the Wilson action of a factor of between 1.9 and 2.4.

However, what is not clear from this is how the saving in iterations translates into the most important quantity, a saving in compute time. Shifting from a standard Wilson action to a partially smeared action means that we now have two sets of gauge fields, the standard and smeared links. This doubles the number of vector-multiplications needed, and the standard spin-projection trick\cite{alford-improving} is no longer applicable, providing an additional factor of 2 in both the multiplications needed and the communications needed. So, moving from the Wilson action to the FLIC action costs us a factor of 4 in vector multiplications and a factor of 2 in communications, plus the overhead for the clover term. On the other hand, evaluating the action of $\epsilon_N(x)$ on a vector costs $O(2N)$ vector multiplications in addition to the two evaluations of the kernel, $H$. While vector multiplications form a significant part of the cost of evaluating $H$, they are not the only part. There is a relatively high cost of communication compared to computation on the parallel architectures that we wish to use. It quickly becomes clear that the only real way to see how much of an improvement we have made is to do an actual calculation and compare the compute time needed.

To test the actual speedup, we choose to calculate the low-lying eigenmodes of $H_{\rm o}^2=D_{\rm o}^\dagger D_{\rm o}$ for the two different kernels, Wilson and FLIC. This calculation allows us to verify that both kernels give the appropriate spectral properties\cite{edwards-practical}, and also allows us to calculate directly the relative compute time needed to evaluate $D_{\rm o}$ in each case. For the Wilson action we used the 14th order Rational Polynomial Approximation in the region which it is bounded by unity $(0.025 < |x| < 1.918)$ and where the maximal deviation from $\epsilon(x)$ is $3.1 \times 10^{-5}$. We used the mass parameter $m=1.65$ and projected out 15 eigenvalues. For the FLIC action, we can take advantage of the improved condition number without reducing the accuracy of our approximation by using the polar decomposition at 12th order, which is sufficient to provide a maximal deviation of less than $3.1 \times 10^{-5}$. This saves us a (small) amount of computation. To optimise the condition number we choose to perform only 6 APE sweeps with the mass parameter set to $m=1.45$ and projecting out 10 eigenvalues. To minimise the computation needed, we implement individual pole convergence testing in our Multi-CG routine. The first pole is considered converged in the $n^{th}$ iteration according to the usual criterion based on the residue, $||r_n|| < \delta$, where we chose $\delta=10^{-8}$. The convergence criterion for the other poles is easily deduced by noting the shifted polynomial structure of the residual, $r^i_n = P_n(H^2+\sigma(i))r_0 = \zeta_n^\sigma P_n(H^2)r_0 = \zeta_n^\sigma r_n$. Then the $i^{th}$ pole is considered converged if
\eqn{ ||r_n||\zeta^{\sigma(i)}_n < 0.1\times\delta, }
where $\zeta^{\sigma(i)}_n$ is defined as in Eq.\ (2.44) of Ref.\ \cite{jegerlehner-multicg}. We have tested this convergence criterion by calculating individual residues and found it to be numerically very safe, and also to save significant amounts of computation. We consider the ten $8^3\times 16, \beta = 4.38$ lattices. Computations are performed on 4 nodes of the Orion supercomputer, (a Sun E420R cluster comprised of 40 nodes, with each node posessing 4 GB of RAM, 16 MB of L2-cache, and 4 UltraSPARC II 450 MHz processors and with nodes are connected by Myrinet networking. The lowest 6 eigenmodes of $H_{\rm o}^2$ are calculated on each configuration using the Ritz functional method \cite{simma-acccg}. We measure the compute time spent in each of the different parts of the ``inner-CG'' calculation, with the following results.

\begin{table}[!htbp]
\begin{center}
\begin{tabular}{lcc}
Code portion                       & Wilson & FLIC6 \\
\hline
1 Kernel-vector evaluation (H)     & 0.022 sec & 0.037 sec \\
1 Multi-CG iteration (including H) & 0.133 sec & 0.154 sec \\ 
1 Multi-CG iteration (excluding H) & 0.089 sec & 0.079 sec \\
1 overlap-vector evaluation        & 25.52 sec & 13.67 sec \\
\end{tabular}
\end{center}
\caption{\label{tab:computetime} Actual compute time spent in the various parts of the algorithm.}
\end{table}

The results show that using the FLIC action as the kernel in the overlap formalism provides a saving of a factor of 1.9 in actual compute-time spent in evaluating the overlap action. This is easily understood by first observing that the time spent in the fermion matrix multiplication constitutes less than half of the compute time spent in the inner CG inversion. Secondly, we have only paid a factor of 2 in compute time moving from the Wilson action to the FLIC action, not the potential factor of 4. This is because the time spent in communication and performing the $\gamma$ matrix algebra is not negligible when compared to the time spent in performing the gauge field multiplications. Finally, as the improved condition number of the FLIC kernel allows us to use the 12th order polar decomposition, we expend less effort per iteration in the CG component of the sign function evaluation. This is because the number of unconverged poles per iteration is reduced, as demonstrated in Table \ref{tab:poleconv}. 

\begin{table}[!htb]
\begin{center}
\begin{tabular}{ccc|ccc}
Pole  & Wilson & FLIC6 & Pole  & Wilson & FLIC6 \\
\hline
 1 & $188_{-21}^{+32}$ & $ 85_{-6}^{+11}$ & 8 & $ 55_{-3}^{+4}$ & $ 19_{-1}^{+1}$ \\
 2 & $188_{-21}^{+32}$ & $ 82_{-4}^{+10}$ & 9 & $ 39_{-2}^{+2}$ & $ 14_{-1}^{+1}$ \\
 3 & $188_{-21}^{+32}$ & $ 65_{-4}^{+6}$ & 10 & $ 28_{-2}^{+1}$ & $ 10_{-0}^{+1}$ \\
 4 & $188_{-21}^{+31}$ & $ 50_{-2}^{+4}$ & 11 & $ 19_{-1}^{+2}$ & $  7_{-1}^{+0}$ \\
 5 & $161_{-13}^{+15}$ & $ 39_{-2}^{+3}$ & 12 & $ 14_{-1}^{+0}$ & $  4_{-0}^{+0}$ \\
 6 & $116_{-8}^{+7}$ & $ 31_{-2}^{+2}$ & 13 & $  9_{-1}^{+0}$ &  - \\
 7 & $ 80_{-5}^{+5}$ & $ 24_{-1}^{+2}$ & 14 & $  5_{-0}^{+1}$ &  - \\
\end{tabular}
\end{center}
\caption{\label{tab:poleconv} Breakdown of the mean convergence for each of the poles.}
\end{table}

These facts mean that the overall compute time per inner CG iteration increases by only 15\% when moving to the FLIC kernel, and hence the saving of 55\% in the total number of inner CG iterations needed translates into a saving in compute time. Thus we have shown that the FLIC action is numerically superior to the Wilson action as an overlap kernel. What has not been answered is what, if any, are the differences in physical properties of $D_{\rm o}$ using the different kernels. For example, overlap fermions are free of $O(a)$ errors irrespective of the choice of kernel, but in general may have different $O(a^2)$ errors. This will be addressed in future work.  

\section{Conclusion}

Practical implementations of the overlap-Dirac operator use a sum over poles to approximate the matrix sign function. These approximations are evaluated using an iterative conjugate gradient routine. As each iteration requires about twice as much computational effort to evaluate as a single evaluation of $H_{\rm w}$, reducing the number of iterations needed is the most direct way of reducing the expense of the overlap formalism. To succeed in this, we select an overlap kernel with an improved condition number motivated by analytic arguments. From the six candidate actions tested, the FLIC action has the best convergence properties, requiring less low-lying projections than the Wilson action and providing a saving in iterations by about a factor of 2. This saving in iterations translates almost directly into a saving in computation time. We restate that only the irrelevant operators are smeared, and that minimal smearing is required, 6 sweeps at $\alpha=0.7$ for $\beta=4.38, a=0.165(2)$ or 4 sweeps at $\alpha=0.7$ for $\beta=4.60, a=0.125(2)$. As the FLIC action has only nearest neighbour couplings, it is well suited to calculations on highly parallel machines. We recognise that there will be some implementation dependence in our compute-time results, but believe that this dependence will be sufficently small that all groups who wish to perform overlap calculations will benefit in moving from the Wilson to the FLIC kernel. As we have concluded that the FLIC action is a numerically superior kernel, we can proceed to investigate the dependence of the overlap action's physical properties on the kernel action. 

\section*{Acknowledgements}

WK wishes to sincerely thank Urs Heller for his correspondence regarding the overlap formalism and also thanks Jianbo Zhang for a number of helpful discussions. We have also benefited from discussions with Robert Edwards, Urs Heller and Herbert Neuberger. Additionally, we thank F. Bonnet for his contribution to the APE smearing code. The calculations were carried out on the Orion Supercomputer of the Australian National Computing Facility for Lattice Gauge Theory. This work was supported by the Australian Research Council.







\end{document}

%% file: commands.tex
\usepackage{amsmath}
\usepackage{amssymb}
\usepackage{amsfonts}
\usepackage{amsthm}

\newcommand{\four}{\!\!\!\!/}

\newtheorem{theorem}{Theorem}[section]
\newtheorem{corollary}[theorem]{Corollary}
\newtheorem{lemma}{Lemma}[section]
\newtheorem{definition}{Definition}[section]
\newtheorem{proposition}{Proposition}[section]
\newtheorem{example}{Example}[section]

\newcommand{\eqn}[1]{ \begin{equation} #1 \end{equation} }

%% file: overlap.bbl
\begin{thebibliography}{40}

\bibitem{overlap1}
R.~Narayanan and H.~Neuberger,
\newblock Phys. Lett. {\bf B302}, 62 (1993), hep-lat/9212019.

\bibitem{overlap2}
R.~Narayanan and H.~Neuberger,
\newblock Nucl. Phys. {\bf B412}, 574 (1994), hep-lat/9307006.

\bibitem{overlap3}
R.~Narayanan and H.~Neuberger,
\newblock Phys. Rev. Lett. {\bf 71}, 3251 (1993), hep-lat/9308011.

\bibitem{overlap4}
R.~Narayanan and H.~Neuberger,
\newblock Nucl. Phys. {\bf B443}, 305 (1995), hep-th/9411108.

\bibitem{neuberger-massless}
H.~Neuberger,
\newblock Phys. Lett. {\bf B417}, 141 (1998), hep-lat/9707022.

\bibitem{wilson}
K.~G. Wilson,
\newblock Phys. Rev. {\bf D10}, 2445 (1974).

\bibitem{neuberger-overlap}
H.~Neuberger,
\newblock Chin. J. Phys. {\bf { 38}}, 533 (2000), hep-lat/9911022.

\bibitem{neuberger-moremassless}
H.~Neuberger,
\newblock Phys. Lett. {\bf B427}, 353 (1998), hep-lat/9801031.

\bibitem{ginsparg-wilson}
P.~H. Ginsparg and K.~G. Wilson,
\newblock Phys. Rev. {\bf D25}, 2649 (1982).

\bibitem{luscher-chiral}
M.~Luscher,
\newblock Phys. Lett. {\bf { B428}}, 342 (1998), hep-lat/9802011.

\bibitem{laliena}
P.~Hasenfratz, V.~Laliena, and F.~Niedermayer,
\newblock Phys. Lett. {\bf B427}, 125 (1998), hep-lat/9801021.

\bibitem{Hasenfratz(NPB)}
P.~Hasenfratz,
\newblock Nucl. Phys. {\bf B525}, 401 (1998), hep-lat/9802007.

\bibitem{Shailesh}
S.~Chandrasekharan,
\newblock Phys. Rev. {\bf D60}, 074503 (1999), hep-lat/9805015.

\bibitem{Giusti}
L.~Giusti, G.~C. Rossi, M.~Testa, and G.~Veneziano,
\newblock (2001), hep-lat/0108009.

\bibitem{neuberger-practical}
H.~Neuberger,
\newblock Phys. Rev. Lett. {\bf { 81}}, 4060 (1998), hep-lat/9806025.

\bibitem{edwards-practical}
R.~G. Edwards, U.~M. Heller, and R.~Narayanan,
\newblock Nucl. Phys. {\bf { B540}}, 457 (1999), hep-lat/9807017.

\bibitem{jegerlehner-multicg}
B.~Jegerlehner,
\newblock (1996), hep-lat/9612014.

\bibitem{neuberger-bounds}
H.~Neuberger,
\newblock Phys. Rev. {\bf { D61}}, 085015 (2000), hep-lat/9911004.

\bibitem{edwards-chiral}
R.~G. Edwards, U.~M. Heller, and R.~Narayanan,
\newblock (1999), hep-lat/9905028.

\bibitem{luscher-review}
M.~Luscher,
\newblock (1998), hep-lat/9802029.

\bibitem{sheik-clover}
B.~Sheikholeslami and R.~Wohlert,
\newblock Nucl. Phys. {\bf B259}, 572 (1985).

\bibitem{gattringer-clover}
C.~Gattringer and I.~Hip,
\newblock Nucl. Phys. {\bf B541}, 305 (1999), hep-lat/9806032.

\bibitem{DeGrand}
T.~DeGrand, A.~Hasenfratz, and T.~G. Kovacs,
\newblock Nucl. Phys. {\bf B547}, 259 (1999), hep-lat/9810061.

\bibitem{stephenson}
M.~Stephenson, C.~DeTar, T.~DeGrand, and A.~Hasenfratz,
\newblock Phys. Rev. {\bf D63}, 034501 (2001), hep-lat/9910023.

\bibitem{ape-one}
M.~Falcioni, M.~L. Paciello, G.~Parisi, and B.~Taglienti,
\newblock Nucl. Phys. {\bf B251}, 624 (1985).

\bibitem{ape-two}
APE, M.~Albanese {\em et~al.},
\newblock Phys. Lett. {\bf B192}, 163 (1987).

\bibitem{derek-apesmearing}
F.~D.~R. Bonnet, P.~Fitzhenry, D.~B. Leinweber, M.~R. Stanford, and A.~G.
  Williams,
\newblock Phys. Rev. {\bf D62}, 094509 (2000), hep-lat/0001018.

\bibitem{derek-smooth}
F.~D.~R. Bonnet, D.~B. Leinweber, A.~G. Williams, and J.~M. Zanotti,
\newblock (2001), hep-lat/0106023.

\bibitem{MIT}
M.~C. Chu, J.~M. Grandy, S.~Huang, and J.~W. Negele,
\newblock Phys. Rev. {\bf D49}, 6039 (1994), hep-lat/9312071.

\bibitem{zanotti-hadron}
J.~M. Zanotti {\em et~al.},
\newblock (2001), hep-lat/0110216.

\bibitem{lepage-mfi}
G.~P. Lepage and P.~B. Mackenzie,
\newblock Phys. Rev. {\bf D48}, 2250 (1993), hep-lat/9209022.

\bibitem{luscher-locality}
P.~Hernandez, K.~Jansen, and M.~Luscher,
\newblock Nucl. Phys. {\bf { B552}}, 363 (1999), hep-lat/9808010.

\bibitem{Fujikawa}
K.~Fujikawa,
\newblock Nucl. Phys. {\bf B546}, 480 (1999), hep-th/9811235.

\bibitem{Suzuki}
H.~Suzuki,
\newblock Prog. Theor. Phys. {\bf 102}, 141 (1999), hep-th/9812019.

\bibitem{Adams-axial}
D.~H. Adams,
\newblock (1998), hep-lat/9812003.

\bibitem{Kikukawa}
Y.~Kikukawa and A.~Yamada,
\newblock Phys. Lett. {\bf B448}, 265 (1999), hep-lat/9806013.

\bibitem{adams}
D.~H. Adams,
\newblock J. Math. Phys. {\bf 42}, 5522 (2001), hep-lat/0009026.

\bibitem{Bietenholz-fast}
W.~Bietenholz, I.~Hip, and K.~Schilling,
\newblock Nucl. Phys. Proc. Suppl. {\bf 106}, 829 (2002), hep-lat/0111027.

\bibitem{Bietenholz-GW}
W.~Bietenholz,
\newblock Eur. Phys. J. {\bf C6}, 537 (1999), hep-lat/9803023.

\bibitem{Hasenfratz-NPPS}
P.~Hasenfratz, S.~Hauswirth, K.~Holland, T.~Jorg, and F.~Niedermayer,
\newblock Nucl. Phys. Proc. Suppl. {\bf 106}, 799 (2002), hep-lat/0109004.

\bibitem{Hasenfratz-IJMP}
P.~Hasenfratz {\em et~al.},
\newblock Int. J. Mod. Phys. {\bf C12}, 691 (2001), hep-lat/0003013.

\bibitem{DeGrand(overlap)}
MILC, T.~DeGrand,
\newblock Phys. Rev. {\bf D63}, 034503 (2001), hep-lat/0007046.

\bibitem{chiu-axial}
T.-W. Chiu and T.-H. Hsieh,
\newblock (2001), hep-lat/0109016.

\bibitem{chiu-index}
T.-W. Chiu,
\newblock Phys. Lett. {\bf B521}, 429 (2001), hep-lat/0106012.

\bibitem{edwards-spectral}
R.~G. Edwards, U.~M. Heller, and R.~Narayanan,
\newblock Nucl. Phys. {\bf { B535}}, 403 (1998), hep-lat/9802016.

\bibitem{alford-improving}
M.~G. Alford, T.~R. Klassen, and G.~P. Lepage,
\newblock Nucl. Phys. {\bf B496}, 377 (1997), hep-lat/9611010.

\bibitem{simma-acccg}
T.~Kalkreuter and H.~Simma,
\newblock Comput. Phys. Commun. {\bf 93}, 33 (1996), hep-lat/9507023.

\end{thebibliography}
